\documentclass[letterpaper, 10 pt, conference]{ieeeconf}  % Comment this line out if you need a4paper

\usepackage{amsmath,amssymb}

\usepackage{amsthm}
\usepackage{custommath}
\usepackage{tikz}

\usepackage{hyperref}
\usetikzlibrary{cd, decorations.pathreplacing}

\hypersetup{colorlinks=true, linkcolor=blue}

\allowdisplaybreaks{}
\newtheorem{cor}{Corollary}

\newtheorem{lem}{Lemma}

\newtheorem{thm}{Theorem}

\IEEEoverridecommandlockouts                              % This command is only needed if 
\title{\LARGE \bf
Matching Observations to Distributions: Efficient Estimation via Sparsified Hungarian Algorithm
}

\author{Sinho Chewi$^1$\thanks{$^1$\texttt{chewisinho@berkeley.edu}}
\and Forest Yang$^2$ \thanks{$^2$\texttt{forestyang@berkeley.edu}}
\and Avishek Ghosh$^3$ \thanks{$^3$\texttt{avishek\_ghosh@berkeley.edu}}
\and Abhay Parekh$^4$ \thanks{$^4$\texttt{parekh@berkeley.edu}}
\and Kannan Ramchandran$^5$ \\ \thanks{$^5$\texttt{kannanr@berkeley.edu}}
}

\begin{document}

\maketitle
\thispagestyle{empty}
\pagestyle{empty}

%%%%%%%%%%%%%%%%%%%%%%%%%%%%%%%%%%%%%%%%%%%%%%%%%%%%%%%%%%%%%%%%%%%%%%%%%%%%%%%%
\begin{abstract}
Suppose we are given observations, where each observation is drawn independently from one of $k$ known distributions.
The goal is to \textit{match} each observation to the distribution from which it was drawn.
We observe that the maximum likelihood estimator (MLE) for this problem can be computed using weighted bipartite matching, even when $n$, the number of observations per distribution, exceeds one. This is achieved by instantiating $n$ duplicates of each distribution node. However, in the regime where the number of observations per distribution is much larger than the number of distributions, the Hungarian matching algorithm for computing the weighted bipartite matching requires $\mathcal O(n^3)$ time. 
We introduce a novel randomized matching algorithm that reduces the runtime to $\tilde{\mathcal O}(n^2)$ by sparsifying the original graph, returning the exact MLE with high probability. 
Next, we give statistical justification for using the MLE by bounding the excess risk of the MLE, where the loss is defined as the negative log-likelihood. We test these bounds for the case of isotropic Gaussians with equal covariances and whose means are separated by a distance $\eta$, and find (1) that $\gg \log k$ separation suffices to drive the proportion of mismatches of the MLE to 0, and (2) that the expected fraction of mismatched observations goes to zero at rate $\mathcal O({(\log k)}^2/\eta^2)$.
\end{abstract}

%%%%%%%%%%%%%%%%%%%%%%%%%%%%%%%%%%%%%%%%%%%%%%%%%%%%%%%%%%%%%%%%%%%%%%%%%%%%%%%%
\section{Introduction}

Suppose we are given observations, where each observation is drawn independently from one of a (finite) set of distributions. The set of distributions as well as the number of observations drawn from each distribution are assumed to be known beforehand. The goal is to \textit{match} each observation to the distribution from which it was drawn. An abstract example is: given $N=nk$ signals, and $k$ sources which each produced $n$ signals, match each signal to the source it came from. More generally, the number of samples coming from each distribution may vary from one another.

%\textbf{[Discussion of related work/more motivation? Differentiate from clustering, which does not constrain number of points per cluster?]}

A natural estimator for the true matching is the maximum likelihood estimator (MLE). 
The MLE may be computed efficiently via weighted bipartite matching, which we now briefly describe.
Given a weighted bipartite graph with $|V|$ vertices and $|E|$ edges, the Hungarian algorithm  runs in time $\mathcal O(|V||E|+|V|^2\log |V|)$ and outputs a maximum weight perfect matching (\cite{fredman1987}).
To compute the MLE, we construct a bipartite graph in which the observations are represented as left nodes, distributions as right nodes, and log probabilities are edge weights.
Since the Hungarian algorithm returns a perfect matching, we require the bipartite graph to be \emph{balanced} (the number of left nodes and right nodes must be equal), so we duplicate each distribution node along with its edges $n$ times to produce a complete bipartite graph with $2nk$ vertices. On this graph, the Hungarian algorithm correctly finds the MLE in $\mathcal O(k^3 n^3)$ time.

We will be interested in the regime where the number of observations tends to infinity while the number of distributions remains constant. In this regime, we give a simple, significantly faster sparsified Hungarian algorithm for computing the MLE exactly with high probability. For each observation node, we only include $c\log n$ randomly chosen edges to each set of distribution nodes, where $c$ is a constant. The number of edges is reduced to $ck^2 n\log n$, so the Hungarian algorithm finds the maximum weight matching on this graph in $\mathcal O(k^3 n^2 \log n + k^2 n^2 \log(kn))$ time. With high probability, the maximum weight matching on the sparsified graph still corresponds to the MLE, which guarantees correctness of our algorithm, while the running time of the algorithm improves from $\mathcal O(n^3)$ to $\tilde{\mathcal O}(n^2)$.

Next, we give a statistical analysis of the MLE for this problem by casting the problem in the framework of \emph{empirical risk minimization}~\cite{koltchinskii2011oracle} and using established tools from empirical process theory. Our first risk bound,~\autoref{thm:excess_risk}, makes no additional assumptions on the distributions from which the data is drawn. The drawback of this bound is that it does not fully utilize \emph{separation} between the data distributions, which can aid the matching. To remedy this, we provide another risk bound,~\autoref{thm:sepwhp}, which is sharper when the distributions are well-separated.

To illustrate our bounds, we consider a simple Gaussian example where the data distributions are standard Gaussian distributions in $\R^k$ with identity covariances and whose mean vectors are the standard basis vectors scaled by a factor $\eta > 0$ (representing the level of separation between the distributions). We show that separation $\eta \gg \log k$ is sufficient for the expected fraction of mismatched observations to tend to zero with $k\to\infty$. In particular, suppose that the number of observations $n$ per distribution does not grow super-polynomially in $k$, i.e., $n \lesssim k^\alpha$ for some $\alpha > 0$, and the separation satisfies $\eta \asymp k^\beta$ for some $0 < \beta \le 1/2$.
Then, our first risk bound shows that the expected fraction of mismatched observations is $\tilde{\mathcal O}(1/k^\beta)$, and our second risk bound improves this to $\tilde{\mathcal O}(1/k^{2\beta})$.

We now briefly compare our separation criteria with the separation requirement in standard clustering problem with Gaussian mixtures.
Our setting differs from $k$-means clustering in two important ways: (1) we assume that the component distributions are known, and (2) our example only considers isotropic Gaussians for the purpose of illustrating our general bounds.
Thus, the problem we study is easier (we do not need to learn the unknown parameters  simultaneously as we cluster the data), and more general (our analysis to more general mixture distributions; in particular, the data need not necessarily take values in Euclidean space).
With that said, for the $k$-means analysis in~\cite{kumar2010clustering, awasthi2012improved, lu2016statistical}, the required separation condition is $\Omega(\sqrt{k})$. Hence, for the special case of a known isotropic Gaussian mixture, our analysis provides a stronger conclusion than the $k$-means analysis.

\subsection{Organization of the Paper}

In Section~\ref{scn:setup}, we formally introduce the relevant definitions and notation for the matching problem. In Subsection~\ref{sscn:review_hungarian}, we briefly review the Hungarian algorithm, which we use to compute the MLE in Subsection~\ref{sscn:mle_graph}.
For readers who are familiar with the Hungarian algorithm, Subsection~\ref{sscn:review_hungarian} can be safely skipped.
In Subsection~\ref{sscn:alg}, we describe our sparsified algorithm and state our main guarantee: the sparsified algorithm computes the MLE exactly, with high probability, using time complexity of only $\tilde{\mathcal O}(n^2)$ rather than the $\mathcal O(n^3)$ time required from a more na\"{\i}ve application of the Hungarian algorithm.
We prove this guarantee in Subsection~\ref{sscn:alg_analysis}.

We state our main statistical guarantees in Subsection~\ref{sscn:risk_bounds}, and test our bounds on an illustrative example with Gaussian samples in Subsection~\ref{sscn:example}. We defer the proofs of all of the bounds to Subsection~\ref{sscn:proofs}.

\subsection{Notation}

We use the notation $a \lesssim b$ or $b \gtrsim a$ to denote that $a \le Cb$ for some universal constant $C > 0$, and we write $a \asymp b$ when both $a \lesssim b$ and $a \gtrsim b$ hold. For a positive integer $m$, $[m]$ denotes $\{1,\ldots, m\}$.
It will be convenient to introduce the notation $\overline Y := Y-\E Y$ for a random variable $Y$.

\section{Matching Observations to Distributions}\label{scn:setup}
\subsection{Setting \& Notation}

Suppose we are given observations, where each observation is drawn from a known (finite) set of probability distributions.
The goal is to \textit{match} each observation to the distribution from which it was drawn.

To formally state this problem in mathematical language, let $k$ be a positive integer, representing the number of categories.
Associated with each category $j \in [k]$ is a probability distribution $P_j$, and $n$ i.i.d.\ observations drawn from $P_j$.
Let $N := nk$ be the total number of observations.
We will be interested in the regime where $k$ is a constant and the number of observations tends to infinity, i.e., $n\to\infty$.
However, our statistical results will be given as finite-sample guarantees.

Let $X_1,\dotsc,X_N$ denote the observed data.
Specifically, since the matching from each observation to the distribution from which it was drawn is initially unknown, we model this as an unknown matching function $\theta^* : [N] \to [k]$ such that for each $i \in [N]$, independently, $X_i \sim P_{\theta^*(i)}$.
We will work in the classical setting of parametric statistics.
Thinking of $\theta^*$ as an unknown parameter, we denote the distribution of $(X_1,\dotsc,X_N)$ by $P_{\theta^*}$, and we set as our parameter space $\Theta$ the set of all matchings $\theta : [N] \to [k]$ which match exactly $n$ observations to category $j$ for each $j\in [k]$, that is, $\abs{\theta^{-1}(\{j\})} = n$ for all $j \in [k]$.
Associated with each $\theta \in \Theta$ is a distribution $P_\theta$ such that if $(X_1,\dotsc,X_N) \sim P_\theta$, then $X_i \sim P_{\theta(i)}$ independently for all $i \in [N]$.

We will focus our study on the \textit{maximum likelihood estimator (MLE)}, defined as follows.
We assume that the distributions $P_1,\dotsc,P_k$ have respective densities $p_1,\dotsc,p_k$ with respect to a common dominating measure.
The \textit{log-likelihood function} for this statistical model is $\ell(\theta; X) := \sum_{i=1}^N \ln p_{\theta(i)}(X_i)$ and the MLE is $\hat \theta := \argmax_{\theta \in \Theta} \ell(\theta; X)$.

\section{A Sparsified Hungarian Algorithm for Computing the MLE}

%\textcolor{blue}{AG comment: May be compress this section (since these are no our contribution and standard stuff), and make a new section titled proofs, where all the proofs will be moved? It is often a good practice to include informal statements of the results in the introduction. So do think about that.  Also, I would switch the statistical guarantee and Sparsified Hungarian algo. I will leave it to you. I think it is better to show guarantees of an estimator and then give an efficient way to compute the estimator.}
\subsection{Review of the Hungarian Algorithm}\label{sscn:review_hungarian}

To recap: given a weighted bipartite graph $G = (V, E, w)$, where $V$ is the disjoint union of $L$ and $R$, and $w_{i,j}$ is the weight of edge $(i, j)$, $i \in L$, $j \in R$, the \emph{maximum weighted bipartite matching} in $G$ is a set of edges $M \subseteq E$ such that $M$ is a \emph{matching} (each vertex is the endpoint of at most one edge in $M$) and $M$ maximizes the sum of weights $\sum_{e \in M} w_e$.
The maximum weighted bipartite matching has an linear programming (LP) relaxation
\begin{align*}
    \begin{aligned}
        & \operatorname*{maximize}_{(x_e, \; e \in E)} && \sum_{e \in E} x_e w_e \\
        & \text{subject to} && \sum_{j \in V : (i,j) \in E} x_{i,j} = 1, & \forall i \in V \\
        &&& x_e \ge 0, & \forall e \in E
    \end{aligned}
\end{align*}
which has a corresponding dual LP
\begin{align*}
    \begin{aligned}
        & \operatorname*{minimize}_{(p_i, \; i \in V)} && \sum_{i \in V} p_i \\
        & \text{subject to} && p_i + p_j \ge w_{i,j}, & \forall (i,j) \in E.
    \end{aligned}
\end{align*}
The dual LP problem is the \emph{weighted vertex cover} problem: given a weighted bipartite graph, assign a \emph{price} to each vertex such that the weight of every edge is at most the sum of the prices of the edge's endpoints, such that the total sum of prices is minimized.
The value of any feasible solution to the dual LP upper bounds the value of any feasible solution to the primal LP\@.

According to classical LP duality theory (see~\cite[\S 6.3]{bertsekas2003convex}), an optimal primal-dual solution satisfies the \emph{complementary slackness conditions for LPs}; in particular, for all $(i, j) \in E$, if $x^*_{i,j} > 0$, then $p^*_i + p^*_j = w_{i,j}$ for a pair of optimal solutions $(x^*_e, \; e \in E)$ and $(p^*_i, \; i \in V)$.

It is well-known that the classical Hungarian algorithm computes the maximum weighted bipartite matching in $\mathcal O(|V|(|E| + |V| \log |V|))$ time (\cite{fredman1987}).
The Hungarian algorithm is a primal-dual algorithm which maintains a feasible weighted vertex cover while iteratively increasing the cardinality of a matching which obeys the complementary slackness conditions.
We next give a brief description of the Hungarian algorithm.

Given a weighted vertex cover $(p_i, \; i \in V)$, call the edge $(i, j) \in E$ \emph{tight} (with respect to the vertex cover) if $p_i + p_j = w_{i,j}$.
Start with a feasible weighted vertex cover by assigning the price $\max_{j \in V : (i, j) \in E} w_{i,j}$ to each $i \in L$, and the price $0$ to each $i \in R$.
Then, create an auxiliary bipartite graph formed from the original bipartite graph by retaining only the edges which are tight with respect to the weighted vertex cover.
Note that if a maximum cardinality matching is found in the auxiliary graph, then the weight of the matching equals the weight of the vertex cover (by the definition of tightness), which provides a certificate of optimality for the primal-dual pair of solutions (the weighted matching and the weighted vertex cover).

To increase the cardinality of a matching in the auxiliary graph, search for an \emph{augmenting path}, that is, a path in the auxiliary graph starting and ending at unmatched vertices such that the edges in the path alternate between unmatched and matched edges (a \emph{matched} edge is an edge present in the current matching).
Augmenting paths can be found using breadth-first search, and the existence of an augmenting path implies that the cardinality of the matching can be increased.

If no augmenting paths can be found, then the set of vertices which can be reached (via paths which alternate between unmatched and matched edges) from unmatched left vertices defines a cut in the graph.
Let $\delta$ be the minimum value of $p_i + p_j - w_{i,j}$ for any edge $(i, j)$ between a left vertex in the cut and a right vertex not in the cut.
For each vertex reachable from the unmatched left vertices, decrease the price of the vertex by $\delta$ if the vertex is a left vertex, and increase the price of the vertex by $\delta$ if the vertex is a right vertex.
This operation is guaranteed to introduce a new tight edge into the auxiliary graph and decrease the overall sum of prices in the weighted vertex cover, while maintaining the dual feasibility of the weighted vertex cover.
The algorithm then continues to find augmenting paths, until the algorithm terminates with an optimal weighted matching and weighted vertex cover whose values coincide.

For further details, we refer readers to~\cite[\S 11.2]{papadimitriou}.
We also remark that the linear programming relaxation of the weighted bipartite matching problem has an integer solution, which provides an alternative polynomial-time algorithm (see~\cite[\S 9.8.3]{moore2011computation}).
For the matching problem, the bipartite graph is complete, so the runtime is $\mathcal O(N^3)$.

\subsection{Computing the MLE via Graph Matching}\label{sscn:mle_graph}

Recall that $k$ and $n$ are positive integers, where $k$ is the number of distributions and $n$ is the number of observations to be matched to each distribution, so that $N = kn$.
Thus, our observation list is indexed by $L = (1,\dotsc, N)$, our distribution list is indexed by $R=(1,\ldots,k)$ and we wish to find the maximum likelihood assignment of $n$ observations to each distribution. 

This is solved by the general problem where one is given a bipartite graph $G=(L,R,E)$ and a weight function $w : E \to \mathbb{R}$ 
such that $|L|=N= kn$ and $|R|=k$. A subset of edges $M$ is called an \emph{$n:1$ matching} if each left node is incident to exactly one edge in $M$ and each right node is incident to exactly $n$ edges in $M$.
The \emph{value} of the matching $M$ is $\sum_{e \in M} w(e)$, and the goal is to find an $n:1$ matching of maximum value. 
Observe that if we associate each edge $(i, j)$ with the weight $\log p_j(X_i)$, then the maximum weight matching recovers the MLE\@. Any $n:1$ matching $M$ corresponds to the function $\theta\in\Theta$ defined by $(i, \theta(i)) \in M$ for all $i\in [N]$, and vice versa. Thus, the $\theta^*$ corresponding to the optimal $M^*$ satisfies 
\begin{equation*}
\theta^* = \argmax_{\theta\in\Theta}
\sum_{i=1}^N w_{i,\theta(i)}
= \argmax_{\theta\in\Theta}\sum_{i=1}^N \log p_{\theta(i)}(X_i).
\end{equation*}

Our initial solution is to form the bipartite graph $G'=(L, R', E')$, where $L$ is the original set of left nodes and $R'$ and $E'$ consist of each node in $R$ along with its edges in the original graph duplicated $n$ times, resulting in a bipartite graph with $kn$ left nodes and $kn$ right nodes.
By duplicating all of the edges, a maximum weighted perfect matching in the augmented graph corresponds to an optimal $n:1$ matching for the original bipartite graph. 
However, application of the Hungarian algorithm to the augmented graph (a complete bipartite graph with $kn$ left nodes and $kn$ right nodes) takes $\Omega(|L||E'|)=\Omega(k^3n^3)$ time.

\subsection{The Proposed Algorithm}\label{sscn:alg}

We form the augmented weighted bipartite graph $G^* = (L, R', E^*)$ where $R'$ consists of $k$ copies each of the $n$ right nodes in $R$.
We use the term \textit{group} to refer to a set of copies of a right node in $R$; thus, $R'$ consists of $k$ groups, each containing $n$ vertices.
An edge $(i, j')$ in $E^*$, where $j'$ is a copy of the vertex $j \in R$, inherits the weight $w(i, j)$ from $G$.

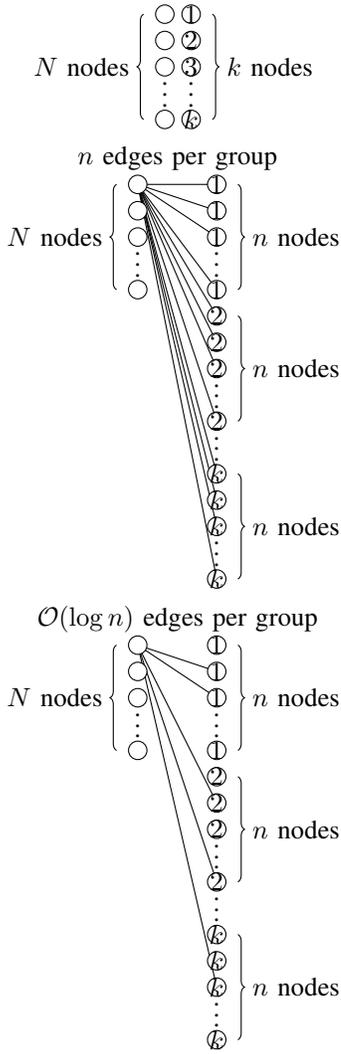
\begin{figure}[!htb]
    \centering
    \begin{tikzpicture}[scale=0.35]
        \draw (0, 0) circle (0.35cm);
        \draw (0, -1) circle (0.35cm);
        \draw (0, -2) circle (0.35cm);
        \node at (0, -2.75) {$\vdots$};
        \draw (0, -4) circle (0.35cm);
        \draw[decoration={brace,mirror,raise=8pt},decorate] (0, 0) -- node[left=10pt] {$N\text{ nodes}$} (0, -4);
        \draw (1, 0) circle (0.35cm) node {$1$};
        \draw (1, -1) circle (0.35cm) node {$2$};
        \draw (1, -2) circle (0.35cm) node {$3$};
        \node at (1, -2.75) {$\vdots$};
        \draw (1, -4) circle (0.35cm) node {$k$};
        \draw[decoration={brace,raise=8pt},decorate] (1, 0) -- node[right=10pt] {$k\text{ nodes}$} (1, -4);
    \end{tikzpicture}
    \linebreak
    \begin{tikzpicture}[scale=0.35]
        \draw (0, 0) -- (3, 0);
        \draw (0, 0) -- (3, -1);
        \draw (0, 0) -- (3, -2);
        \draw (0, 0) -- (3, -4);
        \draw (0, 0) -- (3, -5);
        \draw (0, 0) -- (3, -6);
        \draw (0, 0) -- (3, -7);
        \draw (0, 0) -- (3, -9);
        \draw (0, 0) -- (3, -11);
        \draw (0, 0) -- (3, -12);
        \draw (0, 0) -- (3, -13);
        \draw (0, 0) -- (3, -15);
        \draw[fill=white] (0, 0) circle (0.35cm);
        \draw[fill=white] (0, -1) circle (0.35cm);
        \draw[fill=white] (0, -2) circle (0.35cm);
        \node at (0, -2.75) {$\vdots$};
        \draw[fill=white] (0, -4) circle (0.35cm);
        \draw[decoration={brace,mirror,raise=8pt},decorate] (0, 0) -- node[left=10pt] {$N\text{ nodes}$} (0, -4);
        \draw[fill=white] (3, 0) circle (0.35cm) node {$1$};
        \draw[fill=white] (3, -1) circle (0.35cm) node {$1$};
        \draw[fill=white] (3, -2) circle (0.35cm) node {$1$};
        \node at (3, -2.75) {$\vdots$};
        \draw[fill=white] (3, -4) circle (0.35cm) node {$1$};
        \draw[fill=white] (3, -5) circle (0.35cm) node {$2$};
        \draw[fill=white] (3, -6) circle (0.35cm) node {$2$};
        \draw[fill=white] (3, -7) circle (0.35cm) node {$2$};
        \node at (3, -7.75) {$\vdots$};
        \draw[fill=white] (3, -9) circle (0.35cm) node {$2$};
        \node at (3, -9.75) {$\vdots$};
        \draw[fill=white] (3, -11) circle (0.35cm) node {$k$};
        \draw[fill=white] (3, -12) circle (0.35cm) node {$k$};
        \draw[fill=white] (3, -13) circle (0.35cm) node {$k$};
        \node at (3, -13.75) {$\vdots$};
        \draw[fill=white] (3, -15) circle (0.35cm) node {$k$};
        \draw[decoration={brace,raise=8pt},decorate] (3, 0) -- node[right=10pt] {$n\text{ nodes}$} (3, -4);
        \draw[decoration={brace,raise=8pt},decorate] (3, -5) -- node[right=10pt] {$n\text{ nodes}$} (3, -9);
        \draw[decoration={brace,raise=8pt},decorate] (3, -11) -- node[right=10pt] {$n\text{ nodes}$} (3, -15);
        \node at (1.5, 1) {$n \text{ edges per group}$};
    \end{tikzpicture}
    \begin{tikzpicture}[scale=0.35]
        \draw (0, 0) -- (3, -1);
        \draw (0, 0) -- (3, -2);
        \draw (0, 0) -- (3, -6);
        \draw (0, 0) -- (3, -9);
        \draw (0, 0) -- (3, -13);
        \draw[fill=white] (0, 0) circle (0.35cm);
        \draw[fill=white] (0, -1) circle (0.35cm);
        \draw[fill=white] (0, -2) circle (0.35cm);
        \node at (0, -2.75) {$\vdots$};
        \draw[fill=white] (0, -4) circle (0.35cm);
        \draw[decoration={brace,mirror,raise=8pt},decorate] (0, 0) -- node[left=10pt] {$N\text{ nodes}$} (0, -4);
        \draw[fill=white] (3, 0) circle (0.35cm) node {$1$};
        \draw[fill=white] (3, -1) circle (0.35cm) node {$1$};
        \draw[fill=white] (3, -2) circle (0.35cm) node {$1$};
        \node at (3, -2.75) {$\vdots$};
        \draw[fill=white] (3, -4) circle (0.35cm) node {$1$};
        \draw[fill=white] (3, -5) circle (0.35cm) node {$2$};
        \draw[fill=white] (3, -6) circle (0.35cm) node {$2$};
        \draw[fill=white] (3, -7) circle (0.35cm) node {$2$};
        \node at (3, -7.75) {$\vdots$};
        \draw[fill=white] (3, -9) circle (0.35cm) node {$2$};
        \node at (3, -9.75) {$\vdots$};
        \draw[fill=white] (3, -11) circle (0.35cm) node {$k$};
        \draw[fill=white] (3, -12) circle (0.35cm) node {$k$};
        \draw[fill=white] (3, -13) circle (0.35cm) node {$k$};
        \node at (3, -13.75) {$\vdots$};
        \draw[fill=white] (3, -15) circle (0.35cm) node {$k$};
        \draw[decoration={brace,raise=8pt},decorate] (3, 0) -- node[right=10pt] {$n\text{ nodes}$} (3, -4);
        \draw[decoration={brace,raise=8pt},decorate] (3, -5) -- node[right=10pt] {$n\text{ nodes}$} (3, -9);
        \draw[decoration={brace,raise=8pt},decorate] (3, -11) -- node[right=10pt] {$n\text{ nodes}$} (3, -15);
        \node at (1.5, 1) {$\mathcal O(\log n) \text{ edges per group}$};
    \end{tikzpicture}
    \caption{(Top) The original bipartite graph has $N$ left nodes and $k$ right nodes. (Middle) In the na\"{\i}ve approach, each of the $k$ right nodes is duplicated $n$ times to form a complete bipartite graph with $N$ left nodes and $N$ right nodes. In the figure, only the edges connected to the topmost left node are shown. (Bottom) After sparsification, each left node is only connected to $\mathcal O(\log n)$ right nodes for each of the $n$ groups.}\label{fig:sparsified}
\end{figure}

The idea of the algorithm is very simple.
We construct $E^*$ in the following way: for each vertex $i \in L$ and each group in $R'$ independently, add $c \log n$ edges chosen without replacement, where $c$ is a constant to be chosen later.
This produces a graph with $\mathcal O(k^2n \log n)$ edges, and we run the Hungarian algorithm on the augmented graph, so the time complexity is $\mathcal O(k^3 n^2 \log n)$ as advertised.
See Figure~\ref{fig:sparsified} for a visual depiction of the sparsified graph.
The next theorem establishes the correctness of the algorithm, and it is proved in the next subsection.

\begin{thm}
    For a universal constant $c > 0$, if we add $c\log n$ edges independently from each left node to each right group (chosen uniformly without replacement), then the Hungarian algorithm returns the maximum weighted perfect matching w.h.p.
\end{thm}

\subsection{Analysis of the Algorithm}\label{sscn:alg_analysis}
The correctness of the algorithm hinges on the following:
\begin{quote}
    \textbf{Observation}: If all of the edges of any optimal $n:1$ matching in the original bipartite graph are included in the augmented bipartite graph, then the perfect matching recovered by the Hungarian algorithm will correspond to an optimal assignment in the original bipartite graph.
\end{quote}

In light of the observation, it suffices to bound the probability that no optimal $n:1$ matchings in the original bipartite graph are present in the augmented bipartite graph.
The key is the following:

\begin{thm}[Hall's Marriage Theorem, {\cite[Theorem 5.1]{MR1871828}}]\label{thm:hall}
    Let $G = (L, R, E)$ be a bipartite graph on $n$ left nodes and $n$ right nodes, and for any subset $A \subseteq L$, let $\Gamma(A)$ denote the set of neighbors of vertices in $A$.
    A necessary and sufficient condition for $G$ to have a perfect matching is that for every subset $A \subseteq L$, $\abs A \le \abs{\Gamma(A)}$.
\end{thm}

Fix an optimal $n:1$ matching $M$ in $G$.
We say that $M$ \textit{survives} in $G^*$ if there exists a perfect matching $M'$ in $G^*$ such that for every $(i, j) \in M$, $i$ is matched to a copy of $j$ in $M'$.
In order for $M$ to survive in $G^*$, it is necessary and sufficient that for each group $U\subseteq R'$, if $\{v_1,\dotsc,v_n\} \subseteq L$ is the set of left nodes which are matched (in $M$) to the right node corresponding to the group $U$, then the subgraph of $G^*$ induced by $\{v_1,\dotsc,v_n\}$ and $U$ has a perfect matching.
Therefore, our goal is to prove the following:

\begin{thm}\label{thm:bipartite-augment}
    Let $G = (L, R, E)$ be a bipartite graph with $\abs L = \abs R = n$, such that each vertex in $L$ is connected independently to a uniformly random subset of $c \ln n$ vertices in $R$.
    Then, for a universal constant $c$, $G$ has a perfect matching with high probability.
    
\end{thm}
% \vspace{-2mm}
\begin{proof}
    We can bound the probability that a matching $M$ does not survive in $G^*$ by taking a union bound over the groups $U$ in $R'$ on the event that there does not exist a perfect matching between $U$ and the vertices in $L$ which are matched by $M$ to the vertex corresponding to $U$. The probability of the latter is just the probability that a perfect matching exists in a bipartite graph of $n$ vertices where each left vertex has edges to $c\ln n$ random right vertices.  By Hall's Marriage Theorem (Theorem~\ref{thm:hall}), 
    \begin{align*}
        p_n
        &:= \Pr( \exists A \subseteq L, \; \abs A > \abs{\Gamma(A)} ) \\
        &\le \sum_{k=c\ln n + 1}^n \Pr( \exists A \subseteq L, \; k = \abs A > \abs{\Gamma(A)} ) \\
        %&\le \sum_{k=c\ln n + 1}^n \binom{n}{k} \Pr(\text{a specific } A \subseteq L \text{ with } k \text{ vertices has } \abs{\Gamma(A)} < k) \\
        %&\le \sum_{k=c \ln n + 1}^{n-1} \binom{n}{k}^2 \Pr(\text{a specific } A \subseteq L \text{ only has neighbors in } B \subseteq R, \; \abs A = k = \abs B) \\
        %&\qquad +n\Pr(\Gamma(L) \subseteq B \subseteq R, \; |B|=n-1)\\
        &\le \sum_{k=c\ln n + 1}^{n-1} \binom{n}{k}^2 {\biggl[\frac{\binom{k}{c\ln n}}{\binom{n}{c\ln n}}\biggr]}^k
        +n {\biggl[\frac{\binom{n-1}{c\ln n}}{\binom{n}{c\ln n}}\biggr]}^{n} \\
        &\le \sum_{k=c\ln n + 1}^{n-1} \binom{n}{k}^2 {\Bigl( \frac{k}{n} \Bigr)}^{ck\ln n}
        +n {\Bigl(\frac{n-1}{n}\Bigr)}^{cn\ln n}.
    \end{align*}
    To control the summation, we split the summation into two parts.
    %In the following calculations, $C$ denotes a universal constant that is allowed to change from step to step.
    \begin{align*}
        &\sum_{k=c\ln n + 1}^{\lfloor n/2 \rfloor} \binom{n}{k}^2 {\Bigl( \frac{k}{n} \Bigr)}^{ck \ln n} \\
        &\qquad \le \sum_{k=c\ln n+1}^{\lfloor n/2 \rfloor} {\Bigl( \frac{e n}{k} \Bigr)}^{2k} {\Bigl( \frac{k}{n} \Bigr)}^{ck \ln n} \\
        &\qquad \le \sum_{k=c\ln n+1}^{\lfloor n/2 \rfloor} {\Bigl( \frac{k}{n} \Bigr)}^{k(c \ln n - 2)} e^{2k} \\
        &\qquad \le \sum_{k=c\ln n+1}^{\lfloor n/2 \rfloor} {\Bigl( \frac{1}{2} \Bigr)}^{k (c\ln n - 2)} e^{2k} \\
        &\qquad \le \sum_{k=c\ln n + 1}^\infty {\Bigl( \frac{4e^2}{n^{c\ln 2}} \Bigr)}^k
        = \mathcal O(n^{-c})
    \end{align*}
    for $c$, $n$ sufficiently large.
    For the second part of the summation, for $c > 4$,
    \begin{align*}
        & \sum_{k=\lceil n/2 \rceil}^{n-1} \binom{n}{k}^2 {\Bigl( \frac{k}{n} \Bigr)}^{ck\ln n} \\
        &\qquad = \sum_{k=1}^{\lfloor n/2 \rfloor} \binom{n}{k}^2 {\Bigl( \frac{n-k}{n} \Bigr)}^{c(n-k)\ln n} \\
        &\qquad \le \sum_{k=1}^{\lfloor n/2 \rfloor} {\Bigl( \frac{e n}{k} \Bigr)}^{2k} \exp\Bigl\{ - \frac{ck(n-k)\ln n}{n} \Bigr\} \\
        &\qquad \le \sum_{k=1}^{\lfloor n/2 \rfloor} {\Bigl( \frac{e n}{k} \Bigr)}^{2k} \exp\Bigl\{ - \frac{ck \ln n}{2} \Bigr\} \\
        &\qquad \le \sum_{k=1}^{\lfloor n/2 \rfloor} {\Bigl( \frac{e n}{k} \Bigr)}^{2k} n^{-ck/2}
        \le \sum_{k=1}^{\lfloor n/2 \rfloor} {\Bigl( \frac{e^2}{n^{c/2-2}} \Bigr)}^k \\
        &\qquad = \mathcal O(n^{-(c/2-3)}).
        % \vspace{-2mm}
    \end{align*}
    % \vspace{-2mm}
    Finally, the last term is
    \begin{align*}
        n {\Bigl( \frac{n-1}{n} \Bigr)}^{cn \ln n}
        \le n\exp\{-c \ln n\}
        = \mathcal O(n^{-(c-1)}).
    \end{align*}
    % \vspace{-3mm}
    Thus, we get $p_n \to 0$ as $n\to\infty$.
    %In fact, we have $p_n = O(n^{-c'})$ for some universal constant $c'$ which can be made arbitrarily large by increasing $c = \abs{\Gamma(i)}/(\ln n)$ (where $i$ is a left node), the number of edges per left node.
    In fact, if we take $c = 10$, then $p_n = \mathcal O(n^{-2})$.
\end{proof}

The previous result shows that for each left node, if we attach $10 \ln n$ edges to each group in $G^*$, then the probability that any particular group will fail to have a maximum weight perfect matching surviving from $G$ is $ \mathcal O(n^{-2})$, and taking a union bound over the $k$ groups shows that the algorithm recovers the optimal assignment with probability $1 - \mathcal O(n^{-1})$ (and by using $2(r+4) \ln n$ edges to each group for each vertex, we can achieve a probability of $1- \mathcal O(n^{-r})$ for any positive integer $r$).

The intuition of the algorithm is clearly brought out by the following calculation.
In the setting of Theorem~\ref{thm:bipartite-augment}, the number of possible perfect matchings is $n{!}$, and the probability that a particular perfect matching appears in the random graph is ${[(c \ln n)/n]}^n$.
Thus, the expected number of perfect matchings in the graph is $\approx {(ce^{-1} \ln n)}^n$, which grows exponentially with $n$.
In fact, even if we only connect a constant $c$ edges to each group for each vertex, then the expected number of perfect matchings in the graph is $\approx {(c/e)}^n$, so for $c > e$ then the expected number of perfect matchings still goes to $\infty$ exponentially fast.
The choice of connectivity $c \ln n$ is used in the proof to make the probability of error go to $0$, but it is not wasteful: even with a constant number of edges per vertex per group, the complexity of the Hungarian algorithm is still $\mathcal O(k^2 n^2 \log (kn))$.

\section{Statistical Guarantees}

For $\theta \in \Theta$, we will denote the \textit{loss} by \[ L(\theta; X) = -\frac{1}{n}\ell(\theta; X) = - \frac{1}{n} \sum_{i=1}^N \ln p_{\theta(i)}(X_i), \]
the \textit{risk} by $R(\theta) := \E L(\theta; X)$, and the \textit{excess risk} by $R(\theta) - R(\theta^*)$.
Since $\hat \theta$ is the choice of parameter which minimizes the loss, our statistical problem can be placed in the standard framework of \textit{empirical risk minimization (ERM)}~\cite{koltchinskii2011oracle}.

The statistical guarantees of the MLE are well-studied in other contexts, e.g.~\cite[Theorem 9.14]{keener2010statistics},~\cite[Theorem 5.39]{vaart1998asymptotic}. However, these results typically assume continuity of the likelihood function w.r.t.\ the parameter, which does not hold in our setting.

\subsection{Excess Risk Bounds}\label{sscn:risk_bounds}

For $j,j' \in [k]$, let $\tilde X_j \sim P_j$ (independently across $j \in [k]$), and define
\begin{align*}
    \zeta_{j,j'} := \ln p_{j'}(\tilde X_j) - \ln p_j(\tilde X_j).
\end{align*}
Define
% \begin{align}\label{eq:defM}
%     M(x) := \max_{j,j' \in [k], \; j \ne j'} \abs{\ln p_j(x) - \ln p_{j'}(x)}
% \end{align}
% and using the notation $\norm X_{L^2} := \sqrt{\E(X^2)}$, define
\begin{align}\label{eq:defnormM}
    \sigma^2 := \sum_{j=1}^k \E \max_{j' \in [k]} {(\zeta_{j,j'} - \E \zeta_{j,j'})}^2.
\end{align}
Our first result gives a bound in expectation for the excess risk of the MLE\@.

\begin{thm}\label{thm:excess_risk}
    $\E[R(\hat\theta) - R(\theta^*)] \le 4\sigma \sqrt{k\ln k}$.
\end{thm}

Next, we derive an alternative bound which may be more useful when the separation between the distributions is large. In particular, we will impose the \textit{separation assumption}
\begin{equation}\label{eq:sep}
    \min_{j,j' \in [k], \; j \ne j'} D_{\rm KL}(P_j \mmid P_{j'}) \ge \eta^2 > 0.
\end{equation}
Note that if the separation assumption does not hold for some $\eta > 0$, then two of the distributions are identical.

To state our result, we need a definition.
Given a convex, increasing function $\psi : \R_+ \to \R_+$ with $\psi(0) = 0$ and such that $\psi(x) \to \infty$ as $x\to\infty$, the \textit{Orlicz norm} $\norm \cdot_\psi$ associated with $\psi$ is a norm on the space of real-valued random variables $X$ defined on the same probability space with $\norm X_\psi < \infty$, where $\norm X_\psi := \inf\{t > 0 : \E \psi(\abs X/t) \le 1\}$.
The case corresponding to $\psi_1(x) := \exp x - 1$ is called the \textit{subexponential norm}, see~\cite[\S 2.7]{vershynin2018probability}.

We introduce the quantities
\begin{align*}
    \tilde \sigma^2
    &:= \sum_{j=1}^k \max_{j'\in [k]} \var \zeta_{j,j'}
    \le \sigma^2, \\
    \sigma_{\psi_1}
    &:= \bigl\lVert \max_{j,j' \in [k]}\abs{\zeta_{j,j'}-\E\zeta_{j,j'}}\bigr\rVert_{\psi_1}.
\end{align*}

\begin{thm}
\label{thm:sepwhp}
    Suppose that the distributions $P_1,\dotsc,P_k$ satisfy the separation assumption~\eqref{eq:sep}.
    There exists a universal constant $C > 0$ such that the following holds.
    For every $t \ge 0$, with probability at least $1-2\exp(-t)$, the excess risk satisfies
    \begin{align*}
        &R(\hat\theta)-R(\theta^*)
        \le C\max\Bigl\{ \frac{\sigma^2 \log(kn)}{\eta^2}, \tilde \sigma \sqrt{\frac{t}{n}}, \sigma_{\psi_1} \frac{t\log n}{n} \Bigr\}.
    \end{align*}
\end{thm}

We present a corollary which converts the high probability bound of~\autoref{thm:sepwhp} to a bound in expectation. 

\begin{cor}\label{cor:sepexpectation}
Suppose that the distributions $P_1,\ldots, P_k$ satisfy~\eqref{eq:sep}. Then,
\begin{align*}
    &\E[R(\hat\theta) - R(\theta^*)]
    \lesssim \max\Bigl\{\frac{\sigma^2\log(kn)}{\eta^2}, \frac{\tilde\sigma}{\sqrt{n}},\frac{\sigma_{\psi_1} \log n}{n}\Bigr\}.
\end{align*}
\end{cor}

We remark that when combined with the separation assumption~\eqref{eq:sep}, the excess risk bounds of~\autoref{thm:sepwhp} and~\autoref{cor:sepexpectation} also yield bounds on the number of mismatched observations.

\begin{cor}
Suppose that the distributions $P_1,\ldots, P_k$ satisfy~\eqref{eq:sep}. Let $M$ be the total number of mismatched observations, defined as the number of indices $i \in [N]$ for which the MLE $\hat \theta$ disagrees with the true parameter: $M := \abs{\{i\in [N] : \hat \theta(i) \ne \theta^*(i)\}}$.
Then, the average number of mismatches, $M/N$, satisfies the following bounds.
\begin{enumerate}
    \item There exists a universal constant $C > 0$ such that
    \begin{align*}
    \frac{M}{N}
    &\le C\max\Bigl\{\frac{\sigma^2 \log(kn)}{k\eta^4}, \frac{\tilde \sigma}{k\eta^2} \sqrt{\frac{t}{n}}, \frac{\sigma_{\psi_1} t \log n}{k\eta^2 n}\Bigr\}
\end{align*}
with probability at least $1-\exp(-t)$.
\item In expectation,
\begin{align*}
    \E \frac{M}{N}
    &\lesssim \frac{\sigma\sqrt{\log k}}{\eta^2 \sqrt k}, \\
    \E \frac{M}{N}
    &\lesssim \max\Bigl\{\frac{\sigma^2 \log(kn)}{k\eta^4}, \frac{\tilde \sigma}{k\eta^2 \sqrt n}, \frac{\sigma_{\psi_1} \log n}{k\eta^2 n}\Bigr\}.
\end{align*}
\end{enumerate}
\end{cor}

\subsection{An Illustrative Example}\label{sscn:example}

In this section, we test our bound on a simple Gaussian example. For $j \in [k]$, let $P_j$ be the Gaussian distribution in $\R^k$ with mean $\mu_j := \eta e_j$ (a scaling of the $j$th standard basis vector in $\R^k$) and identity covariance matrix. A standard calculation shows that the separation between $P_j$ and $P_{j'}$ for $j\ne j'$ is $\asymp \eta^2 \|e_j - e_{j'}\|_2^2 \asymp \eta^2$, so the separation assumption~\eqref{eq:sep} holds (up to a universal constant factor).

We next calculate the relevant quantities to apply our bounds. Write $Z_1,\dotsc,Z_k$ for i.i.d.\ standard Gaussians in $\R^k$.
Note that $\zeta_{j,j'} = \langle \tilde X_j, \mu_{j'} - \mu_j \rangle$, so
\begin{align*}
    \sigma^2
    &= \sum_{j=1}^k \E\max_{j' \in [k]} {(\zeta_{j,j'} - \E \zeta_{j,j'})}^2 \\
    &= \sum_{j=1}^k \E\max_{j' \in [k]} \langle Z_j, \mu_j - \mu_{j'} \rangle^2
    \lesssim \eta^2 k \log k,
\end{align*}
where we used the fact that by Gaussian concentration, $\max_{j' \in [k]} \langle Z_j, \mu_j - \mu_{j'} \rangle$ is sub-Gaussian with variance proxy $\lesssim \eta^2$ (c.f.~\cite[Lemma 6.12]{vhprobability} for concentration of the maximum, and~\cite[Proposition 2.5.2]{vershynin2018probability} for characterization of sub-Gaussian random variables).
Similarly, we calculate
\begin{align*}
    \tilde \sigma^2
    &:= \sum_{j=1}^k \max_{j'\in [k]} \var \zeta_{j,j'}
    = \sum_{j=1}^k \max_{j' \in [k]} \var \langle Z_j, \mu_j - \mu_{j'} \rangle \\
    &\lesssim \eta^2 k, \\
    \sigma_{\psi_1}
    &:= \bigl\lVert \max_{j,j' \in [k]}\abs{\zeta_{j,j'}-\E\zeta_{j,j'}}\bigr\rVert_{\psi_1} \\
    &= \bigl\lVert \max_{j,j' \in [k]} \abs{\langle Z_j, \mu_j - \mu_{j'} \rangle}\bigr\rVert_{\psi_1}
    \lesssim \eta \sqrt{\log k},
\end{align*}
where the second calculation comes from the fact that the sub-Gaussian norm of the maximum of $k^2$ random variables, each with sub-Gaussian norm $\lesssim \eta$, is $\lesssim \eta \sqrt{\log(k^2)} \asymp \eta \sqrt{\log k}$.
Using these bounds, we see that the expected fraction of mismatched observations satisfies
\begin{align*}
    \E \frac{M}{N}
    &\lesssim \frac{\log k}{\eta}
\end{align*}
based on our first excess risk bound, and
\begin{align*}
    \E \frac{M}{N}
    &\lesssim \frac{\log(kn)\log k}{\eta^2} + \frac{1}{\eta\sqrt{kn}} + \frac{(\log k)(\log n)}{\eta kn} \\
    &\lesssim \frac{\log(kn)\log k}{\eta^2} + \frac{1}{\eta\sqrt{kn}}
\end{align*}
based on our second excess risk bound.

The first bound tells us that $\eta \gg \log k$ suffices for the expected fraction of mismatches to vanish, and the error is $\mathcal O((\log k)/\eta)$.
To exhibit a regime in which the second bound is sharper, suppose that $k$ is growing and that $n$ does not grow super-polynomially in $k$, i.e., $n \lesssim k^\alpha$ for some $\alpha > 0$.
In the regime where the separation does not grow too fast, i.e., $\log k \ll \eta \lesssim \sqrt k$, then the first term in the second bound dominates, and the error decreases at rate $\mathcal O({(\log k)}^2/\eta^2)$. Thus, we see that the second rate captures a faster rate of decay of the error.

% This last bound is intuitively of the correct order, as the following reasoning shows. Suppose that $n=1$ so that we have a single observation from each Gaussian, and we want to decide which observation comes from the first Gaussian $P_1$. Let $X_j \sim P_j$ for $j \in [k]$ denote our observations.
% A natural strategy is to project the observations onto the subspace spanned by $e_1$, since $\E \langle X_1, e_1 \rangle = \eta$ and $\E \langle X_j, e_1 \rangle = 0$ for $j = 2,\dotsc,k$.
% Then, we can choose the observation with the largest projection onto $e_1$.
% The strategy succeeds if $\langle X_1, e_1 \rangle > \max_{j=2,\dotsc,k} \langle X_j, e_1 \rangle$.
% Note that the LHS is of order $\mathcal O(\eta)$ and the RHS, being the maximum of $k-1$ independent standard Gaussians, is of the order $\mathcal O(\sqrt{\log k})$.\footnote{In electrical engineering terms, the \emph{signal-to-noise ratio (SNR)} for this problem is $\mathcal O(\eta/\sqrt{\log k})$.} Thus, we expect that $\eta \gg \sqrt{\log k}$ is necessary and sufficient to correctly match the Gaussian observations in this example, which agrees with our bound.

\subsection{Proofs}\label{sscn:proofs}

\subsubsection{Proof of Excess Risk Bound}

Our analysis will proceed via standard empirical process theory arguments.

%The first step will be to split the data $(X_1,\dotsc,X_N)$ into i.i.d.\ $k$-tuples.
Denote by ${(X'_{i,j})}_{i \in [n], \; j \in [k]}$ the list $(X_1,\dotsc,X_N)$ sorted such that for each $i \in [n]$, $j \in [k]$, we have $X_{i,j}' \sim P_j$; moreover, with a slight abuse of notation, if $X_{i,j}'$ corresponds to the observation $X_k$ in the original list, then we denote $\theta(i,j) := \theta(k)$ for all $\theta \in \Theta$.
Thus, $X_{i,j}' \sim P_{\theta^*(i,j)} = P_j$.
With this new notation, we have $\ell(\theta; X) = \sum_{i=1}^n \sum_{j=1}^k \ln p_{\theta(i,j)}(X'_{i,j})$.
%For $i \in [n]$, write \[ \ell'_{\theta, i}(x_1,\dotsc,x_k) := \sum_{j=1}^k \ln p_{\theta(i,j)}(x_j), \] so $\ell(\theta; X) = \sum_{i=1}^n \ell'_{\theta, i}(X'_{i,1},\dotsc,X'_{i,k})$.
%Thus, associated with each $\theta \in \Theta$ is a collection of functions $\ell'_\theta := (\ell'_{\theta,1},\dotsc, \ell'_{\theta,n})$.

Note that
\begin{align*}
    L(\theta; X)
    &= -\frac{1}{n} \sum_{i=1}^n \sum_{j=1}^k \ln p_{\theta(i,j)}(X_{i,j}'), \\
    R(\theta)
    &= \frac{1}{n} \sum_{i=1}^n \sum_{j=1}^k [D_{\rm KL}(P_j \mmid P_{\theta(i,j)}) + H(P_j)],
\end{align*}
where $H$ is the Shannon entropy and $D_{\rm KL}$ is the relative entropy.
Since $\hat \theta = \argmin_{\theta \in \Theta} L(\theta; X)$, the excess risk of $\hat \theta$ can be controlled via the standard inequality
\begin{align}
    &R(\hat \theta) - R(\theta^*)
    = R(\hat \theta) - L(\hat \theta; X) + L(\theta^*; X) - R(\theta^*) \\
    &\qquad\qquad\qquad\qquad {} +\underbrace{L(\hat \theta; X) - L(\theta^*; X)}_{\le 0} \\
    &\qquad \le \max_{\theta \in \Theta} \abs{L(\theta; X) - L(\theta^*; X) - R(\theta) + R(\theta^*)} \label{eq:basic-ineq} \\
    &\qquad = \max_{\theta \in \Theta} \Bigl\lvert \frac{1}{n} \sum_{i=1}^n \sum_{j=1}^k [\overline{\ln p_{\theta(i,j)}(X_{i,j}')} - \overline{\ln p_j(X_{i,j}')}] \Bigr\rvert.
\end{align}

To control the maximum, we need a slight extension of the standard symmetrization argument~\cite[\S 2.1]{koltchinskii2011oracle}. We prove it for the sake of completeness, although the proof is identical to the usual proof of symmetrization.

\begin{lem}[Symmetrization]\label{lem:symmetrization}
    Let $X_1,\dotsc,X_n$ be independent random variables taking values in a space $\mc X$ and let $\mc F$ be a finite class of $n$-tuples of real-valued functions on $\mc X$.
    Let $\tilde X_1,\dotsc,\tilde X_n$ be an independent copy of $X_1,\dotsc,X_n$ and let $\varepsilon_1,\dotsc,\varepsilon_n$ be i.i.d.\ Rademacher random variables independent of all other random variables.
    Then,
    \begin{align*}
        &\E \max_{(f_1,\dotsc,f_n) \in \mathcal F} \Bigl\lvert \sum_{i=1}^n \overline{f_i(X_i)} \Bigr\rvert
        \le 2\E\max_{(f_1,\dotsc,f_n) \in \mathcal F} \Bigl\lvert \sum_{i=1}^n \varepsilon_i \overline{f_i(X_i)} \Bigr\rvert.
    \end{align*}
\end{lem}
\begin{proof}
    Introduce an i.i.d.\ copy $\tilde X_1,\dotsc,\tilde X_n$ of $X_1,\dotsc,X_n$ which is also independent of the Rademacher random variables (enlargening the probability space if necessary).
    Then, applying Jensen's Inequality,
    \begin{align*}
        &\E \max_{(f_1,\dotsc,f_n) \in \mc F} \Bigl\lvert \sum_{i=1}^n [f_i(X_i) - \E f_i(X_i)] \Bigr\rvert \\
        &\qquad= \E \max_{(f_1,\dotsc,f_n) \in \mc F} \Bigl\lvert \sum_{i=1}^n [f_i(X_i) - \E f_i(\tilde X_i)] \Bigr\rvert \\
        &\qquad \le \E \max_{(f_1,\dotsc,f_n) \in \mc F} \Bigl\lvert \sum_{i=1}^n [f_i(X_i) - f_i(\tilde X_i)] \Bigr\rvert \\
        &\qquad = \E \max_{(f_1,\dotsc,f_n) \in \mc F} \Bigl\lvert \sum_{i=1}^n \varepsilon_i [f_i(X_i) - f_i(\tilde X_i)] \Bigr\rvert \\
        &\qquad = \E \max_{(f_1,\dotsc,f_n) \in \mc F} \Bigl\lvert \sum_{i=1}^n \varepsilon_i [\overline{f_i(X_i)} - \overline{f_i(\tilde X_i)}] \Bigr\rvert \\
        &\qquad \le 2\E\max_{(f_1,\dotsc,f_n) \in \mc F} \Bigl\lvert \sum_{i=1}^n \varepsilon_i \overline{f_i(X_i)} \Bigr\rvert. \qedhere
    \end{align*}
\end{proof}

Applying the symmetrization lemma to our setting, we obtain the bound
\begin{align}
    &R(\hat\theta) - R(\theta^*) \\
    &\qquad \le 2\E \max_{\theta \in \Theta} \Bigl\lvert \frac{1}{n} \sum_{i=1}^n \sum_{j=1}^k \varepsilon_{i,j} \overline{(\ln p_{\theta(i,j)} - \ln p_j)(X_{i,j}')} \Bigr\rvert, \label{eq:symmetrized1}
\end{align}
where ${(\varepsilon_{i,j})}_{i\in [n], \; j \in [k]}$ are i.i.d.\ Rademacher random variables independent of ${(X_{i,j}')}_{i\in [n], \; j \in [k]} = {(X_i)}_{i=1}^N$.

We recall that a random variable $Y$ is sub-Gaussian~\cite{boucheron2013concentration} with variance proxy $\tau^2$ if $\E \exp(\lambda Y) \le \exp(\lambda^2 \tau^2/2)$ for all $\lambda \in \R$.
Conditional on $X_1,\dotsc,X_N$, the random variable $n^{-1/2} \sum_{i=1}^n \sum_{j=1}^k \varepsilon_{i,j} [\overline{\ln p_{\theta(i,j)}(X_{i,j}')} - \overline{\ln p_j(X_{i,j}')}]$ is sub-Gaussian with variance proxy
\begin{align*}
    &\max_{\theta' \in \Theta} \frac{1}{n} \sum_{i=1}^n \sum_{j=1}^k {[\overline{\ln p_{\theta'(i,j)}(X_{i,j}')} - \overline{\ln p_j(X_{i,j}')}]}^2 \\
    &\qquad \le \frac{1}{n} \sum_{i=1}^n \sum_{j=1}^k \max_{j' \in [k]} {[\overline{\ln p_{j'}(X_{i,j}')} - \overline{\ln p_j(X_{i,j}')}]}^2.
\end{align*}
Next we recall the following maximal inequality for sub-Gaussian random variables: if ${(Y_t)}_{t\in T}$ is a finite collection of $\tau^2$-sub-Gaussian random variables, $\abs T > 1$, then \[ \E\max_{t\in T} {\abs{Y_t}} \le 2\sqrt{\tau^2 \ln|T|}, \] see~\cite[Theorem 2.5]{boucheron2013concentration}.
Starting from~\eqref{eq:symmetrized1} and applying the maximal inequality, we bound the excess risk as
%\textcolor{blue}{Forest: should it be $\ln 2k$, since $|T| = 2|\Theta|$ due to the absolute value in \eqref{eq:symmetrized1}?}
\begin{align*}
    &\E[R(\hat \theta) - R(\theta^*)] \\
    &\qquad \le 4\E \sqrt{\frac{N\ln k}{n} \frac{1}{n} \sum_{i=1}^n \sum_{j=1}^k \max_{j' \in [k]} {\overline{(\ln p_{j'} - \ln p_j)(X_{i,j}')}}^2} \\
    &\qquad \le \frac{4\sqrt{N\ln k}}{\sqrt n} \sqrt{\sum_{j=1}^k \E\max_{j'\in [k]} {[\overline{\ln p_{j'}(\tilde X_j)} - \overline{\ln p_j(\tilde X_j)}]}^2} \\
    &\qquad = 4\sqrt{k\ln k} \sqrt{\sum_{j=1}^k \E \max_{j'\in [k]}{\abs{\overline{\ln p_{j'}(\tilde X_j)} - \overline{\ln p_j(\tilde X_j)}}}^2}.
\end{align*}
Recalling the definition~\eqref{eq:defnormM}, we have the result.

\subsubsection{Proof of Excess Risk Bound for Large Separation}

Our method will use a standard \textit{fixed point} argument, as detailed in \cite[\S 1.2]{koltchinskii2011oracle}.
Let $\hat \delta := R(\hat \theta) - R(\theta^*)$ denote the excess risk.
The starting point is that in the inequality~\eqref{eq:basic-ineq}, the maximum can be restricted to
\begin{align}\label{eq:fixed-pt}
    \hat \delta
    &\le \max_{\theta \in \Theta(\hat\delta)} \abs{L(\theta; X) - L(\theta^*; X) - R(\theta) + R(\theta^*)}
\end{align}
where for $\delta > 0$, $\Theta(\delta) := \{\theta \in \Theta : R(\theta) \le R(\theta^*) + \delta\}$.
For $\theta \in \Theta(\delta)$, the separation assumption ensures that
\begin{align*}
    \delta
    &\ge R(\theta)-R(\theta^*)
    = \frac{1}{n} \sum_{i=1}^n \sum_{j=1}^k D_{\rm KL}(P_j \mmid P_{\theta(i,j)}) \\
    &\ge \frac{\eta^2 d_{\rm H}(\theta, \theta^*)}{n}
\end{align*}
where $d_{\rm H}$ is the Hamming distance; thus, $d_{\rm H}(\theta, \theta^*) \le \delta n/\eta^2$.
By an elementary combinatorial argument, the number of $\theta \in \Theta$ which are at Hamming distance $\xi$ from $\theta^*$ is at most $\binom{N}{\xi} \xi! \le N^\xi$, so $\abs{\Theta(\delta)} \le N^{\delta n/\eta^2}$.

To control the RHS of~\eqref{eq:fixed-pt}, we start by controlling the expected maximum for a fixed $\delta > 0$.
Applying symmetrization (\autoref{lem:symmetrization}) and a similar argument as before,
\begin{align}
    &\E \max_{\theta \in \Theta(\delta)} \abs{L(\theta; X) - L(\theta^*; X) - R(\theta) + R(\theta^*)} \\
    &\qquad \le \frac{4\sigma\sqrt{\ln{\abs{\Theta(\delta)}}}}{\sqrt n}
    = 4\sigma \sqrt{\frac{\delta}{\eta^2} \ln N}, \label{eq:sup-bound}
\end{align}
where $\sigma^2$ is the variance proxy as before.

Next, we will apply the Adamczak concentration inequality~\cite{adamczak2008tail}, in the form given in~\cite[\S 2.3]{koltchinskii2011oracle}.

\begin{thm}[Adamczak]
    Let $Y_1,\dotsc,Y_n$ be independent $S$-valued random variables, and let $\mc F$ be a countable class of measurable functions $S \to \R$. Assume that for every $f \in \mc F$ and $i\in [n]$, $\E f(Y_i) = 0$, and for some $\alpha \in (0, 1]$ and all $i \in [n]$, $\norm{\sup_{f \in \mc F} \abs{f(Y_i)}}_{\psi_\alpha} < \infty$.
    Let $Z := \sup_{f \in \mc F} \abs{\sum_{i=1}^n f(Y_i)}$ and $\tau^2 := \sup_{f \in \mc F} \sum_{i=1}^n \E[{f(Y_i)}^2]$.
    Then, there exists $K(\alpha) > 0$ such that for all $t \ge 0$,
    \begin{align}\label{eq:adamczak}
    \begin{aligned}
        &\Pr\bigl\{Z \ge K(\alpha)\bigl[\E Z + \tau\sqrt{t} + \bigl\lVert \max_{i \in [n]} \sup_{f \in \mc F} |f(Y_i)| \bigr\rVert_{\psi_\alpha} t^{1/\alpha} \bigr]\bigr\} \\
        &\qquad \le \exp(-t).
        \end{aligned}
    \end{align}
\end{thm}

To apply this result, let $Y_{i,j} := (X_{i,j}', (i,j))$ for $i\in [n]$, $j\in [k]$ and $\mc F(\delta) := \{f_{\theta} : \theta \in \Theta(\delta)\}$, where for $\theta \in \Theta$,
\begin{align*}
    f_{\theta}\bigl(x,(i,j)\bigr)
    &:= \ln p_{\theta(i,j)}(x) - \ln p_j(x) \\
    &\qquad{} - \E[\ln p_{\theta(i,j)}(\tilde X_j) - \ln p_j(\tilde X_j)].
\end{align*}
Then, the ${(Y_{i,j})}_{i\in [n], \; j \in [k]}$ are independent, with
\begin{align*}
   Z
   &:= \max_{f \in \mc F(\delta)} \Bigl\lvert \sum_{i=1}^n \sum_{j=1}^k f(Y_{i,j}) \Bigr\rvert \\
   &= n\max_{\theta \in \Theta(\delta)} \abs{L(\theta; X) - L(\theta^*; X) - R(\theta) + R(\theta^*)}.
\end{align*}
Then,
\begin{align*}
    \tau^2
    &:= \max_{f \in \mc F(\delta)} \sum_{i=1}^n \sum_{j=1}^k \E[{f(Y_{i,j})}^2] \\
    &= \max_{\theta \in \Theta(\delta)} \sum_{i=1}^n \sum_{j=1}^k \E[{\abs{\overline{\ln p_{\theta(i,j)}(X_{i,j}')} - \overline{\ln p_j(X_{i,j}')}}}^2] \\
    &\le n \sum_{j=1}^k \max_{j'\in [k]} \E[{\abs{\overline{\ln p_{j'}(\tilde X_j)} - \overline{\ln p_j(\tilde X_j)}}}^2]
    = n\tilde \sigma^2.
\end{align*}
Using the fact that the Orlicz norms are increasing, i.e., if $X \ge Y$ then $\norm X_\psi \ge \norm Y_\psi$ for any Orlicz norm $\norm \cdot_\psi$,
\begin{align*}
    &\bigl\lVert \max_{i\in [n]} \max_{j\in [k]} \max_{f \in \mc F(\delta)} \abs{f(Y_{i,j})} \bigr\rVert_{\psi_\alpha} \\
    &\qquad \le \bigl\lVert \max_{i\in [n]} \max_{j\in [k]} \max_{\theta \in \Theta}{\abs{\overline{\ln p_{\theta(i,j)}(X_{i,j}')} - \overline{\ln p_j(X_{i,j}')}}} \bigr\rVert_{\psi_\alpha} \\
    &\qquad \le \bigl\lVert \max_{i\in [n]} \max_{j\in [k]} \max_{j' \in [k]}{\abs{\overline{\ln p_{j'}(X_{i,j}')} - \overline{\ln p_j(X_{i,j}')}}} \bigr\rVert_{\psi_\alpha}.
    %&\qquad = \Bigl\lVert \max_{i\in [n]} \max_{\theta \in \Theta(\delta)} \Bigl\lvert \sum_{j=1}^k \overline{\ln p_{\theta(i,j)}(X_{i,j}')} - \overline{\ln p_j(X_{i,j'})} \Bigr\rvert \Bigr\rVert_{\psi_\alpha} \\
    %&\qquad \le \Bigl\lVert \max_{i\in [n]} \sum_{j=1}^k \max_{j' \in [k]} \abs{\overline{\ln p_{j'}(X_{i,j}')} - \overline{\ln p_j(X_{i,j'})}} \Bigr\rVert_{\psi_\alpha}.
\end{align*}

Now, we focus on $\alpha = 1$ and invoke~\cite[Lemma 2.2.2]{vaart1996empirical}.
%We prove the lemma for completeness, but a more general version of this lemma appears as~\cite[Lemma 2.2.2]{vaart1996empirical}.

% \begin{lem}
%     Let $\norm \cdot_\psi$ be any Orlicz norm.
%     For any non-negative random variable $X$, $\norm{\E X}_\psi \le \norm X_\psi$.
% \end{lem}
% \begin{proof}
%     By convexity of $\psi$, for any $t > \norm X_\psi$,
%     \begin{align*}
%         1
%         &\ge \E\psi\Bigl(\frac{X}{t}\Bigr)
%         \ge \psi\Bigl(\frac{\E X}{t}\Bigr),
%     \end{align*}
%     so $\norm{\E X}_\psi \le t$.
% \end{proof}

\begin{lem}
    Let $Y_1,\dotsc,Y_n$ be random variables with $n\ge 2$.
    Then, $\norm{\max_{i\in [n]} Y_i}_{\psi_1} \lesssim (\log n)\max_{i\in [n]}\norm{Y_i}_{\psi_1}$.
\end{lem}
% \begin{proof}
%     The Orlicz norm $\norm X_{\psi_1}$ equals, up to a constant factor, the smallest positive constant $K > 0$ such that $\Pr\{\abs X \ge t\} \le 2\exp(-t/K)$ for all $t\ge 0$, see~\cite[Proposition 2.7.1]{vershynin2018probability}.
%     Let $K := \max_{i\in [n]} \norm{Y_i}_{\psi_1}$.
%     Using a union bound, there exists a universal constant $C > 0$ such that
%     \begin{align*}
%         \Pr\bigl\{\max_{i\in [n]}\abs{Y_i} \ge t\bigr\}
%         &\le 2n\exp\Bigl(- \frac{t}{CK}\Bigr) \qquad \forall t \ge 0.
%     \end{align*}
%     If $t \le 2CK\ln n$, then
%     \begin{align*}
%         \Pr\bigl\{\max_{i\in [n]}\abs{Y_i} \ge t\bigr\}
%         &\le 1
%         \le 2\exp\Bigl(- \frac{t\ln 2}{2CK\ln n}\Bigr).
%     \end{align*}
%     If $t \ge 2CK\ln n$, then
%     \begin{align*}
%         \Pr\bigl\{\max_{i\in [n]}\abs{Y_i} \ge t\bigr\}
%         &\le \Pr\Bigl\{\max_{i\in [n]} \abs{Y_i} \ge CK\ln n + \frac{t}{2}\Bigr\} \\
%         &\le 2\exp\Bigl(- \frac{t}{2CK}\Bigr).
%     \end{align*}
%     The result now follows from the characterization of the subexponential norm.
% \end{proof}

Note that the lemma does not require independence, but the lemma is typically sharp only for independent random variables.
Applying the lemma,
\begin{align*}
    &\bigl\lVert \max_{i\in [n]} \max_{j\in [k]} \max_{f \in \mc F(\delta)} \abs{f(Y_{i,j})} \bigr\rVert_{\psi_1} \\
    &\qquad \lesssim (\log n) \bigl\lVert \max_{j,j' \in [k]}{\abs{\overline{\ln p_{j'}(\tilde X_j)} - \overline{\ln p_j(\tilde X_j)}}}\bigr\rVert_{\psi_1} \\
    &\qquad = \sigma_{\psi_1} \log n.
    %&\qquad \lesssim (\log n)\Bigl\lVert \sum_{j=1}^k \max_{j' \in [k]} \abs{\overline{\ln p_{j'}(\tilde X_j)} - \overline{\ln p_j(\tilde X_j)}} \Bigr\rVert_{\psi_1} \\
    %&\qquad \le (\log n) \sum_{j=1}^k \bigl\lVert \max_{j' \in [k]} \abs{\overline{\ln p_{j'}(\tilde X_j)} - \overline{\ln p_j(\tilde X_j)}} \bigr\rVert_{\psi_1} \\
    %&\qquad = \sigma_{\psi_1} \log n.
\end{align*}
At this point, we will focus on $\alpha = 1$.
By~\eqref{eq:sup-bound} and~\eqref{eq:adamczak}, we have the concentration bound, valid for all $t \ge 0$:
\begin{align*}
    \frac{Z}{n}
    \lesssim \sigma \sqrt{\frac{\delta}{\eta^2} \log N} &+ \tilde \sigma \sqrt{\frac{t}{n}}
    + \sigma_{\psi_1} \frac{t\log n}{n}
\end{align*}
with probability at least $1-\exp(-t)$.
Using arguments from~\cite{koltchinskii2011oracle}, which are summarized concisely in~\cite[Lemma 5.1]{ahidar2018barycenters}, it follows that $\hat \delta$ is bounded above, with probability at least $1-2\exp(-t)$, by the smallest $a^* > 0$ for which
\begin{align*}
    \sigma \sqrt{\frac{\log N}{\eta^2 a^*}} &+ \Bigl(\tilde \sigma\sqrt{\frac{t}{n}} + \sigma_{\psi_1} \frac{t\log n}{n}\Bigr)\frac{1}{a^*}
    \lesssim 1.
\end{align*}
Hence, with probability at least $1-2\exp(-t)$,
\begin{align*}
    \hat\delta
    &\lesssim \max\Bigl\{ \frac{\sigma^2 \log N}{\eta^2}, \tilde \sigma \sqrt{\frac{t}{n}}, \sigma_{\psi_1} \frac{t\log n}{n} \Bigr\}.
\end{align*}

\begin{proof}[Proof of~\autoref{cor:sepexpectation}]
    Let $C>0$ be as in~\autoref{thm:sepwhp}. Then, define constants $a_1 := C\sigma^2 (\log N)/\eta^2$, $a_2 := C\tilde \sigma/\sqrt{n}$, $a_3 := C\sigma_{\psi_1}\log n/n$, and let $b :=\max\{a_2, a_3\}$. Denote $\hat\delta := R(\hat\theta)-R(\theta^*)$ as the excess risk. 
\begin{align*}
    \E\hat\delta
    &= \int_0^\infty \Pr(\hat\delta > t)\,\D t
    \leq a_1 + \int_0^\infty \Pr(\hat\delta > a_1+t)\,\D t \\
    &= a_1 + b\int_0^\infty \Pr(\hat\delta > a_1 + bt)\,\D t \\
    &\leq a_1 + b + b\int_1^\infty \Pr(\hat\delta > a_1 + bt)
        \, \D t \\
    &\leq a_1 + b + b\int_1^\infty \Pr(\hat\delta > \max\{a_1, 
    a_2\sqrt{t}, a_3t\})\,\D t \\
    &\leq a_1 + b + b\int_1^\infty 2\exp(-t) \, \D t
    \lesssim a_1 + b.
\end{align*}
$\hat\delta$ is non-negative because it equals a sum of KL divergences. The third inequality holds because $a_1+bt \geq \max\{a_1, a_2\sqrt{t}, a_3t\}$, since $a_1, bt$ are non-negative and $t\geq\sqrt{t}$ for $t\geq 1$. The fourth inequality uses~\autoref{thm:sepwhp}. 
\end{proof}

\bibliographystyle{IEEEtran}
\bibliography{IEEEabrv,IEEEexample}

\end{document}